\documentclass{cernrep}
\usepackage{graphicx,here}

\def\be{\begin{equation}}
\def\ee{\end{equation}}

\begin{document}

\title{STUDY OF COINCIDENCES BETWEEN RESONANT 
GRAVITATIONAL WAVE DETECTORS}

\author{
P. Astone$^1$, M. Bassan$^2$, P. Bonifazi$^3$, P. Carelli$^4$,\\
E. Coccia$^2$, C.Cosmelli$^5$, S.D'Antonio$^5$,
V. Fafone$^6$, G.Federici$^1$,\\ A. Marini$^6$,
Y. Minenkov$^2$, I. Modena$^2$,\\
G. Modestino$^6$,
A. Moleti$^2$, G. V. Pallottino$^5$,\\ G. Pizzella$^7$,
L.Quintieri$^6$, F. Ronga$^6$, R. Terenzi$^3$, 
M. Visco$^3$, L. Votano$^6$ \\
$~$\\
$~$
}

\vskip 0.1 in

\institute{
{\it ${}^{1)}$ Istituto Nazionale di Fisica Nucleare INFN, Rome}\\
{\it ${}^{2)}$ University of Rome "Tor Vergata" and INFN, Rome}\\
{\it ${}^{3)}$ IFSI-CNR and INFN, Frascati}\\
{\it ${}^{4)}$ University of L'Aquila and INFN, Rome}\\
{\it ${}^{5)}$ University of Rome "La Sapienza" and INFN, Rome}\\
{\it ${}^{6)}$ Istituto Nazionale di Fisica Nucleare INFN, Frascati}\\
{\it ${}^{7)}$ University of Rome "Tor Vergata" and INFN, Frascati}
}

\maketitle
 
\begin{abstract}
Coincidences are searched with the cryogenic
resonant gravitational wave detectors EXPLORER and NAUTILUS,
during a period of about six months (2 June-14 December 1998)
for a total measuring time of 94.5 days, with the purpose to
study new algorithms of analysis, based on the 
physical characteristics of the detectors.

\end{abstract}

%{\bf~~~~June 2000~~~~For circulation only within the ROG group}

%\pagestyle{plain}
%\setcounter{page}2
%\baselineskip=17pt
\section{Introduction}
After the initial experiments with room temperature resonant detectors, the new
generation of cryogenic gravitational wave (GW) 
antennas entered long term data
taking operation in 1990 (EXPLORER~\cite{long}), in 1991
 (ALLEGRO~\cite{alle}), in 1993 (NIOBE~\cite{NIOBE}),
in 1994 (NAUTILUS~\cite{naut}) and in 1997
(AURIGA~\cite{auri}).

Recently an analysis of the data taken in coincidence among all
cryogenic resonant detectors in operation during the years 1997 and 1998
has been performed~\cite{5barre}.
No coincidence excess was found above background using
the event lists produced under the protocol of the
International Gravitational Event Collaboration (IGEC), among the
groups of ALLEGRO, AURIGA, EXPLORER / NAUTILUS and NIOBE.
The coincidence search was done without any particular data
selection. However one can consider the possibility to search
for coincidences with events selected according to
various possible criteria using all available information
(we mention criteria based on:
the event energy, the event duration, the applied threshold,
the shape of the events, the coincidence window, the direction of possible GW,
the noise).

Here we have used algorithms based on physical
characteristics of the detectors, as the event energy (with a new
algorithm) and the directionality. In this paper we explore their
effect on the coincidence search.

For this purpose we shall use IGEC data obtained from 2 June 1998
when NAUTILUS, after a stop for instrumental improvements,
resumed the operation. We search for coincidences between
NAUTILUS and EXPLORER, whose apparatuses differ only in the operating
temperatures (respectively 0.15 K and 2.6 K) and in particular
have identical readout systems. Extension of the methods we develop
here to other detectors in operation during the same period of time
is envisaged.

We are well aware that any data selection
jeopardizes the possibility to express the results
by means of a $probability$ that a coincidence excess, if any, 
had been accidental.
With this $proviso$ we shall still use parameters obtained from
probability estimations for comparing different situations.

\begin{table}
\centering
\caption{
Main characteristics of the two detectors.
}
\vskip 0.1 in
\begin{tabular}{|c|c|c|c|c|c|c|}
\hline
detector&latitude&longitude&orientation&mass&frequencies&temperature\\
&&&&kg&Hz&K\\
\hline
EXPLORER&46.45 N&6.20 E&$39^o$ E&2270&904.7&2.6\\
&&&&&921.3&\\
NAUTILUS&41.82 N&12.67 E&$44^o$ E&2270&906.97&0.15\\
&&&&&922.46&\\
\hline
\end{tabular}
\label{dire}
\end{table}

\section{Events and signals}
We now briefly describe how we obtain $events$ from the measurements.
For EXPLORER and NAUTILUS,
whose main characteristics are given in table \ref{dire},
the data are sampled at intervals of 4.54 ms
and are filtered with a filter matched to short bursts~\cite{fast}
for the detection of delta-like signals. The filter makes use of power
spectra obtained with off-line analysis.
After the filtering of the raw-data, $events$ are extracted as follows.
Be $x(t)$ the filtered output of the detector.
This quantity is normalized, using the detector calibration,
such that its square gives the energy innovation E of the oscillation
for each sample, expressed in kelvin units.
For well behaved noise due only to the thermal
motion of the bar and to the electronic noise of the amplifier,
the distribution of $x(t)$ is normal with zero mean.
The variance (average value of the square of $x(t)$)
is called $effective~temperature$ and is indicated with $T_{eff}$. The
distribution of $x(t)$ is
\be
f(x)=\frac{1}{\sqrt{2\pi T_{eff}}}e^{-\frac{x^2}{2T_{eff}}}
\label{normal}
\ee
For extracting $events$ we set a 
threshold in terms of a critical ratio defined by
\be
 CR=\frac{|x|-\bar{|x|}}{\sigma(|x|)}=
\frac {\sqrt{SNR}-\sqrt {\frac {2}{\pi}}}{\sqrt {1-\frac {2}{\pi}}}
\label{creq}
\ee
where $\sigma(|x|)$ is the standard deviation of $|x|$ 
(the moving averages $\bar{|x|}$
are made over the preceeding ten minutes) and
\be
 SNR=\frac{E}{T_{eff}}
\label{teffeq}
\ee

The threshold is set at CR=6 in order to obtain,
in presence of thermal and electronic noise alone,
about one hundred $events$ per day, as agreed among the partners of the IGEC.  
 This threshold corresponds to an energy
$E_t=19.5~ T_{eff}$. When $|x|$ goes above the threshold, its
time behaviour is considered until it goes below the threshold for
more than ten seconds. The maximum amplitude and its occurrence time
define the $event$.

In general the $event$ is due to a combination of a signal which,
in absence of noise, has energy $E_s$ (due to
GW or other forces) and the noise.
The theoretical probability to detect a signal with a given
 $SNR_s=\frac{E_s}{T_{eff}}$,
in presence of a well behaved Gaussian noise, is calculated as follows.
We put  $y=(s+x)^2$ where
$s\equiv \sqrt{SNR_s}$ is the signal we look for and $x$ is the gaussian
noise. We obtain easily \cite{papoulis}
\be
probability(SNR_s)=\int_{SNR_t}^{\infty} \frac {1}{\sqrt{2 \pi y}}
e^{-\frac{(SNR_s+y)}{2}}
cosh(\sqrt{y\cdot SNR_s})dy
\label{papou}
\ee
where we put $SNR_t=\frac{E_t}{T_{eff}}=19.5$
 for the present EXPLORER and NAUTILUS detectors.

\begin{figure}
 \vspace{9.0cm}
\includegraphics{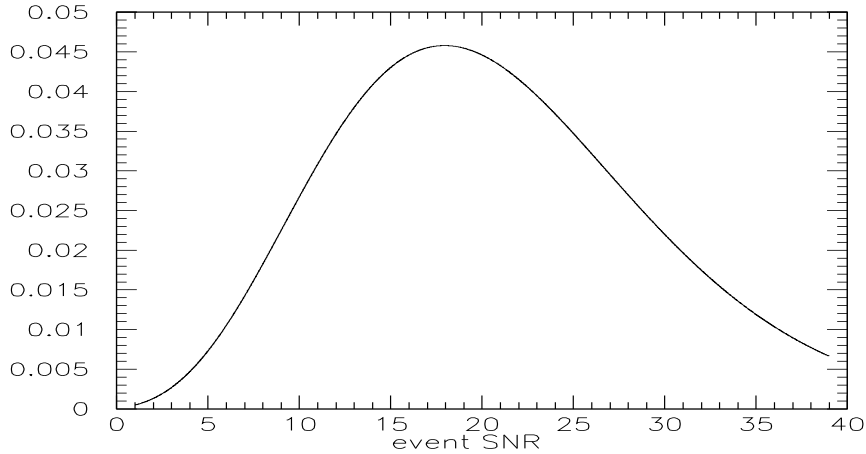}
 \caption{
Differential probability that the event has the signal-to-noise
ratio shown on the abscissa when the signal has $SNR_s=20$.
        \label{snrrappo} }
\end{figure}

The behaviour of the integrand is shown in fig. \ref{snrrappo}.
This figure shows the spread of the event energy due to noise for a given
$SNR_s$ of the applied signal. The distinction between the two concepts,
$signal$ and $event$, is essential for the analysis we propose in this
paper.

\section{Data selection}

All the events which are in coincidence
within a time window of $\pm 5~s$ with events produced by a seismometer
are eliminated, about $8\%$ of the events.

It has been noticed
that the experimental data are affected by noise which, in some cases,
cannot be observed with any other auxiliary detector.
 Thus a strategy is needed for deciding
when the measurements are considered to be good for the search of
coincidences.

We are well aware that the selection of the experimental data must
be done with great care and the safest strategy is to establish rules
before even looking to the data.
We have decided to take into consideration $all$ the data recorded by the
detectors (except those vetoed by the seismometer)
and accept only the events for which the corresponding
$T_{eff}$ is below a certain
threshold. This threshold must be such that we are confident
that no signal is being thrown away.
$All$ and $only$ the events which have
$T_{eff} \leq 100 ~mK$ (over the preceeding ten minutes) are
taken into consideration.
The events for which the corresponding $T_{eff}$ is greater than
 $100 ~mK$ are
certainly generated at times the detector is not operating properly.

The following information is available on the IGEC Web page for each event:\\
Time (UT) of the maximum of the event: YEAR, MONTH, DAY, MINUTE, SECOND.\\ 
$H_o$: Bilateral Fourier amplitude at resonance of the maximum.\\
SNR: Signal to noise ratio of amplitude.\\
$T_{eff}$: Effective temperature [K] of the previous 10 minutes.\\
Duration L of the event, in number of samples (4.54 ms).\\
Time in seconds between the beginning and the maximum of the event.

The relationship between the Fourier transform $H_o$ of the event amplitude
and the energy E of the event is given by \cite{australia}
\be
H_o=7.97~ 10^{-21} \sqrt{E}
\label{acca}
\ee
with $H_o$ in units of $\frac{1}{Hz}$ and E in kelvin.

\begin{figure}
 \vspace{9.0cm}
\includegraphics{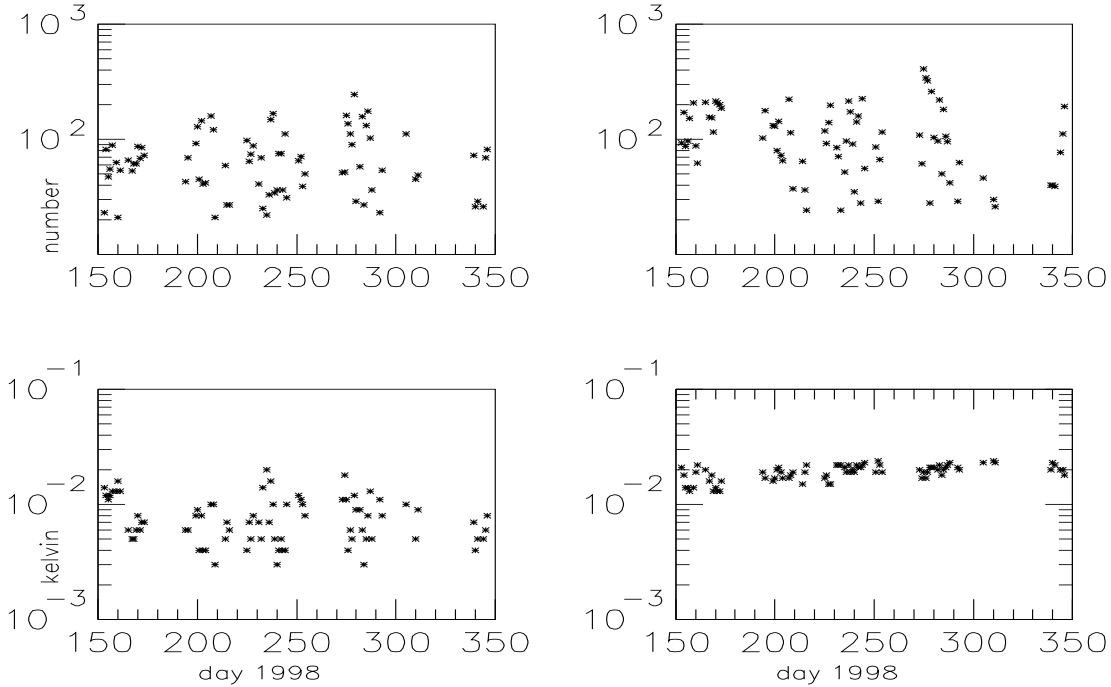}
 \caption{
$T_{eff}\leq25~mK$.
The upper two figures show the number of events/day for
NAUTILUS (left) and EXPLORER (right).
The lower two figures show the noise temperature $T_{eff}$ (kelvin)
respectively for NAUTILUS and EXPLORER, daily averaged over the events.
        \label{stazpuliti} }
\end{figure}

Looking to these events we have noticed that some events occur during
periods of high disturbance. 
Since we are elaborating here strategies for data analysis, we
have thought convenient to select the data
to be used in our analysis in various ways according to the noise.
Thus another way of choosing the data to
be analysed is to select periods with smaller noise. We apply two more
data selections, only events with 
$T_{eff}\leq50~mK$ for both EXPLORER and NAUTILUS
and only events with $T_{eff}\leq25~mK$.
This data selection has been applied by us in a previously
published paper \cite{explniobe}.
In fig.\ref{stazpuliti} we show the number of events
per day and the hourly averages of $T_{eff}$ for EXPLORER and
NAUTILUS for the case $T_{eff}\leq25~mK$.

We notice large fluctuations, in spite of the stringent criteria
for the data selection. Information on the
various data selections are given in table \ref{2criteri}. We notice
that the number of available hours of measurement becomes rather
smaller when lowering the threshold for $T_{eff}$, so that any
possible result becomes statistically weaker at low $T_{eff}$.
\begin{table}
\centering
\caption{
Total number N of events, number of hours of data taking, average
noise temperature $<T_{eff}>$ and
hours in common, when both detectors were simultaneously operating.
}
\vskip 0.1 in
\begin{tabular}{|c|c|c|c|c||c|c|}
\hline
&$T_{eff}$&N&hours&$<T_{eff}>$&hours&N\\
&mK&&&mK&in common&\\
\hline
\hline
EXPLORER&$\leq100$&55070&3415&40.6&2271&37944\\
NAUTILUS&&37734&3450&19.1&&24118\\
\hline
EXPLORER&$\leq50$&39211&2759&28.9&1816&26481\\
NAUTILUS&&34148&3371&14.0&&16677\\
\hline 
EXPLORER&$\leq25$&16172&1498&18.7&931&9765\\
NAUTILUS&&27823&3168&9.3&&5999\\ 
\hline
\end{tabular}
\label{2criteri}
\end{table}

\section{Searching for coincidences}

For the search of coincidences it is important to establish
the time window. We have decided to adopt the same window used in
past analyses, in particular that described in papers \cite{ae1991,5barre},
$w=\pm1~s$. This is a reasonable choice considering the present
detectors bandwidth (of the order of 1 Hz) and some time inaccuracy.

As well known, the analysis in a coincidence search consists essentially
in comparing the detected coincidences at zero time delay 
with the background, that is with coincidences occurring by chance.
In order to measure the background due to the accidental
coincidences, using a procedure adopted since the beginning of
the gravitational wave experiments \cite{weber},
we have shifted the time of occurrence of the events of one of
the two detectors 1,000 times in steps of 2 s, from -1,000 s to
+1,000 s. For each time shift we get a number of coincidences.
If the time shift is zero we get the number $n_c$ of $real$ coincidences.
The background is calculated from the average number of the $n_{shift}$
accidental coincidences obtained from the one thousand time shifts
\be
\bar{n}=\frac{\sum_1^{1000}n_{shift}}{1000}
\label{average}
\ee
With this experimental procedure for the evaluation of the background
we circumvent the problems arising from a non very stationary
distribution of the events, provided we test properly the distribution
of the shifted coincidences (see fig. \ref{delay} and reference \cite{afp}). 

The result of our search for coincidences is given in table \ref{combined}.
\begin{table}
\centering
\caption{
Number $n_c$ of coincidences and average number $\bar{n}$
of accidentals.
The total period of time in common when $T_{eff}\leq100~mK$ is 94.5 days.
}
\vskip 0.1 in
\begin{tabular}{|c|c|c|c|}
\hline
$T_{eff}$&$n_c$&$\bar{n}$&hours\\
\hline
$\leq100~mK$&223&231.7&2271\\
$\leq50~mK$&137&139.8&1816\\
$\leq25~mK$&32&36.2&931\\
\hline
\end{tabular}
\label{combined}
\end{table}
There is no coincidence
excess between EXPLORER and NAUTILUS, even for selected periods
with smaller noise.

\section{Data selection using the event energy}

We want now to apply data selection algorithms based on the
event energy. The most obvious one is to search for pairs of events
which have (approximately) the same energy. In the past this
energy criterion has been applied, requiring that the responses
of the EXPLORER detector at the two resonance modes were within
a factor of two one from each other\cite{long}. Later we realized
that the effect of the noise on
signals near threshold is such that the event energies are
only lightly correlated to the signal energies \cite{snr},
and this reduces the efficiency of algorithms based on
the event energy.

Recently an important result was found \cite{blair}. It has been seen
that the distribution of the energy ratios of the event
energies of two detectors, in the case of non gaussian noise, is different
for real coincidences and accidental coincidences.
This has pushed us to reconsider the importance to apply selection
algorithms based on the event energies.

For making use of the event energy, in particular with detectors
with different sensitivity, we must consider the result
shown in fig. \ref{snrrappo} which indicates the chance to have
a certain event-energy for a given signal-energy. In principle,
all event-energies are possible, from zero to infinity. Our procedure
here is to consider only event-energies within $\pm$ one sigma
from the signal energy (that is, we consider events included
in 68\% of the area under the line in fig. \ref{snrrappo}).

We do not know the signal-energy. The new algorithm we propose 
is the following. We consider signals in a wide range, say: $E_s$ from 20 mK
to 2 K in steps of 20 mK. We find the coincident events, at zero
delay (the real coincidences) and at shifted times (for the estimation
of the accidentals). For each assumed signal with energy $E_s$ we calculate
the $SNR_s$ different for each event, since the noise $T_{eff}$
depends on the detected event and it is also different for
the two detectors.
We then verify if the $SNR_{event}$ falls into $SNR_s \pm1~sigma$,
having calculated for each $SNR_s$ the probability curve
like that shown in fig. \ref{snrrappo} for $SNR_s=20$.
If the two event energies are compatible with the event-energy expected
for any of the assumed signals then we accept the coincidence
(real or shifted).

The result of this analysis is given in table \ref{combi}.
\begin{table}
\centering
\caption{
Energy algorithm.
Number $n_c$ of coincidences, average number $\bar{n}$
of accidentals and the covered time period for the
three data selection.
}
\vskip 0.1 in
\begin{tabular}{|c|c|c|c|}
\hline
$T_{eff}$&$n_c$&$\bar{n}$&hours\\
\hline
$\leq100~mK$&61&50.5&2271\\
$\leq50~mK$&45&37.7&1816\\
$\leq25~mK$&11&10.3&931\\
\hline
\end{tabular}
\label{combi}
\end{table}
We notice that the use of the energy selection algorithm has reduced
the number of the accidental coincidences by a factor of three.

\section{Event selection according to the detector orientation
with respect to the Galactic Centre}

No extragalactic GW signals should be detected with the present
detectors. Therefore we shall focus our attention on possible
sources located in the Galaxy. If any of these sources
exist we should expect a more favorable condition of detection
when the detectors are oriented with their axes perpendicular
to the direction toward the Galactic Centre (GC), since, 
the bar cross-section is proportional to $sin^4(\theta)$,
where $\theta$ is the angle between
the detector axis and the direction to the GC.

After having applied the above energy algorithm,
we search for coincidences considering only events
obtained when the detectors were oriented with $\theta$ greater
than various given values and, according to the previous sections, 
for the various data selection.

\begin{figure}
 \vspace{9.0cm}
\includegraphics{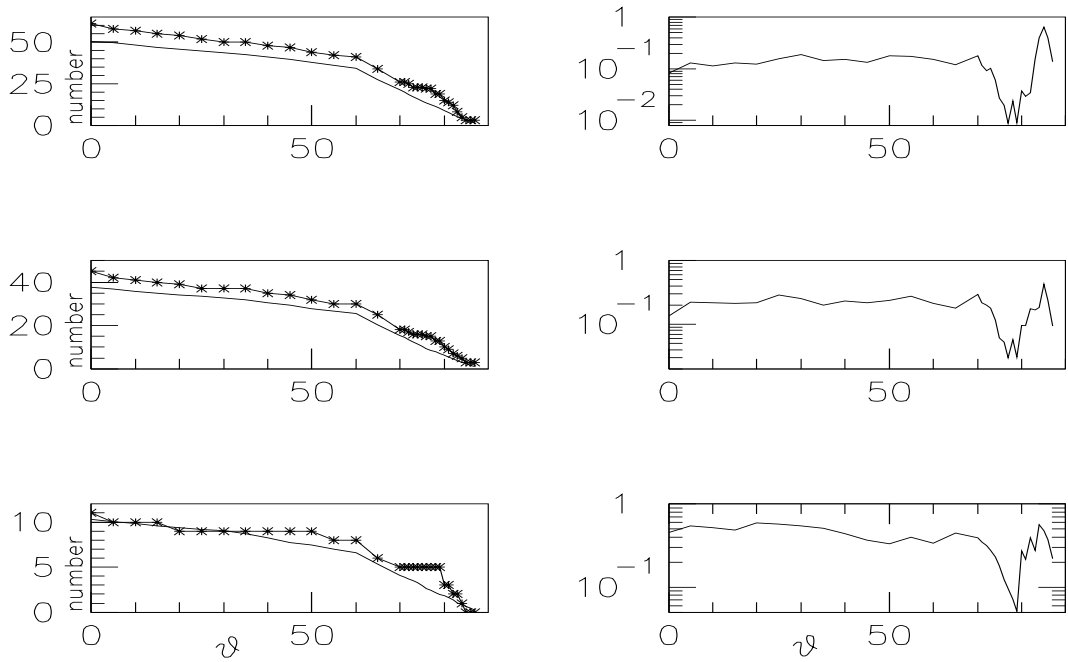}
 \caption{
The upper left figure shows, versus $\theta$ and for $T_{eff}\leq100~mK$,
the integral number of coincidences $n_c$ (indicated with asterisks)
and the average number of accidentals $\bar{n}$
(calculated from
the number of coincidences at zero delay and the average
background $\bar{n}$ measured with 1000 delays).
The right figure shows the Poisson probability that the
observed number of coincidences $n_c$ was due to a
background fluctuation.
Similarly the second line of figures refers to the data selection
$T_{eff}\leq50~mK$ and the third line to the data selection
$T_{eff}\leq25~mK$.
        \label{versoteta} }
\end{figure}

The result is given in fig.\ref{versoteta}
and shows a larger coincidence excess when the detector
axes tend to be perpendicular to the direction towards the GC.
Above $\theta\sim79^o$
the number $n_c$ of coincidences drops quickly.
If not instrumental, the quick drop
could be taken as due to the width of the source. The time spent by
the detectors when $\theta\geq79^o$ is 20\% of the total time
of 94.5 days.

\begin{figure}
 \vspace{9.0cm}
\includegraphics{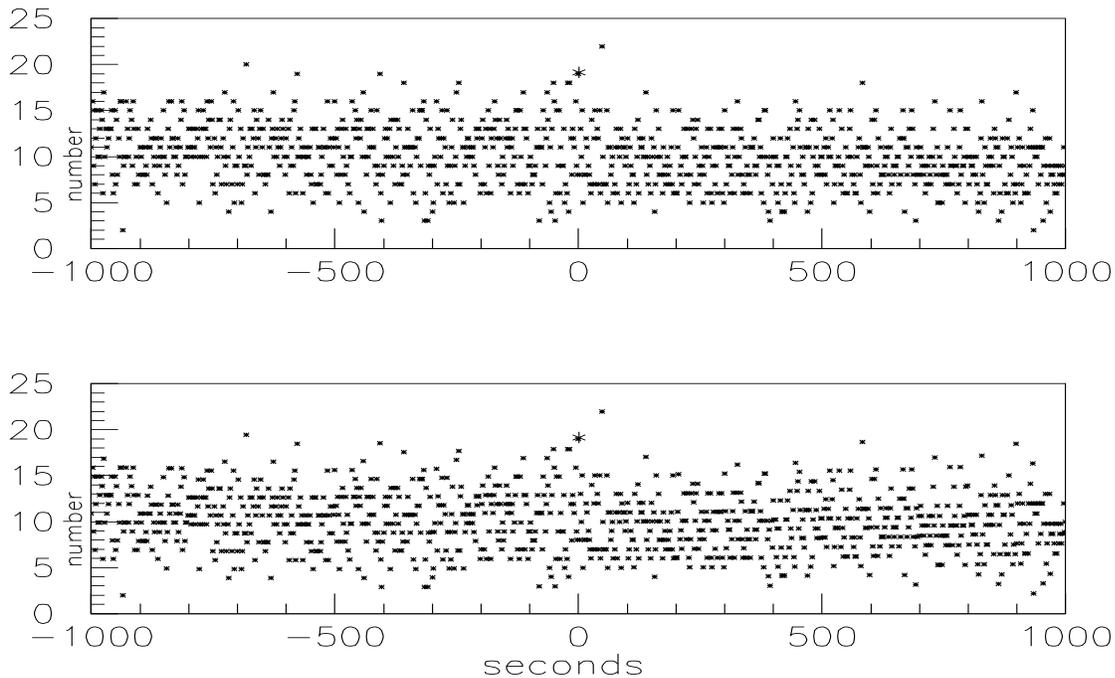}
 \caption{
Data selection with $T_{eff}\leq100~mK$.
In the upper figure we show, for $\theta\geq79^o$, the delay histogram.
In the lower figure we show the same data
normalized for the duration of the time period used for each time
shift. In this particular case the normalization turns out to be
very small, almost barely visible for delays above 700 s, but it is worth
to remark that a possible effect due to different time coverage at
various delays has been taken into account.
 The biggest asterisk indicate the nineteen coincidences at
zero time delay.
        \label{delay} }
\end{figure}

We want now to verify that the evaluation of the background
is properly done. We do this
in the condition of the greatest coincidence excess, that is for
$\theta\geq79^o$. We must consider
that by selecting only times when the detectors had certain
orientations we have several empty time regions. This makes it
possible that in doing the shifting operation for evaluating the
background one uses time periods of different duration. We have determined
these time periods and found that they vary by a few percent,
with a maximum of -10\% for a time shift of +1000 s.
In fig. \ref{delay} we show, for the case $\theta\geq79^o$
and $T_{eff}\leq100~mK$,
the delay histogram with no correction and the delay
histogram corrected for the different periods of time for each shift.
In this particular case the correction applies for a very small amount
only for delays greater than about 700 s.
\section{Conclusions}

In order to make a first step to a complete analysis, we have selected the
IGEC events of EXPLORER and NAUTILUS using algorithms based on known physical
characteristics of the detectors. In particular a new algorithm 
which makes use of the event energy has been devised.

With event selection based on this algorithm we find an excess of coincidences
at zero time delay in the direction of the Galactic Centre. As well known
in the scientific community, no g.w. signals are expected to be observed
with the present detector sensitivity. Since this result would open new
possibilities, a careful Bayesan approach suggests that, given the
Poisson probabilities of a few percent, new data with other detectors
are required before we can ensure that g.w. from the GC have been
indeed observed. Thus, at present, we feel that the coincidence
excess is not large enough to establish a claim
for detection of true signals, but it is an important information
to make available to the scientific community.
We believe that the procedures adopted here might be useful for
detecting gravitational waves with more or better data.

\section{Acknowledgements}
We thank W.O.Hamilton and W.W.Johnson for discussions and suggestions.
We thank the European Center for Nuclear Physics (CERN) for the hospitality
and for the supply of the cryogenic liquids. We thank W.O.Hamilton
and W.W.Johnson for useful suggestions and
 F. Campolungo, R. Lenci, G. Martinelli, E. Serrani, R. Simonetti 
and F. Tabacchioni for precious technical assistance.

%
%\section{References}
%

%
\end{document}